\documentclass[twocolumn,showpacs,preprintnumbers,amsmath,amssymb]{revtex4}
\usepackage[utf8]{inputenc}
\usepackage{epsf}
\usepackage{graphicx}
\usepackage{latexsym}
\usepackage{amsmath}
\usepackage{amssymb}
\usepackage{amsfonts}
\usepackage{color}

\begin{document}


\title{Tunneling interferometry and measurement of thickness of \\ ultrathin metallic Pb(111) films}
\author{S.\,S. Ustavschikov$^{a,b}$, A.\,V. Putilov$^a$ and A.\,Yu. Aladyshkin$^{a,b}$}
\affiliation{$^{a}$ Institute for Physics of Microstructures RAS, 603950, Nizhny Novgorod, GSP-105, Russia \\
$^{b}$ N.\,I. Lobachevsky State University of Nizhny Novgorod, Nizhny Novgorod, 603950 Russia}


\begin{abstract}
Spectra of the differential tunneling conductivity for ultrathin lead films grown on Si(111)$7\times7$ single crystals with a thickness from 9 to 50 monolayers have been studied by low--temperature scanning tunneling microscopy and spectroscopy. The presence of local maxima of the tunneling conductivity is typical for such systems. The energies of maxima of the differential conductivity are determined by the spectrum of quantum--confined states of electrons in a metallic layer and, consequently, the local thickness of the layer. It has been shown that features of the microstructure of substrates, such as steps of monatomic height, structural defects, and inclusions of other materials covered with a lead layer, can be visualized by bias--modulation scanning tunneling spectroscopy.
\end{abstract}



\maketitle

\section{Introduction}

The  main  trend  in  the development of modern solid--state electronics is to reduce dimensions of logical elements, sensors, and conductors  connecting them \cite{Moore}. A consequence of a transition from  macro- to nanoscale is an increase in the influence of quantum confinement effects and effects of discreteness of electric charge and structural disorder on the transport properties of nanoelectronic devices \cite{FG-book}.

In this work we discuss the diagnostics of the quality of deposited metallic layers and the presence of foreign inclusions on the example of ultrathin Pb films. Such films appear to be convenient objects for studying quantum--size effects in normal and superconducting metal films \cite{Altfeder-PRL-1997,Altfeder-PRL-2002,Su-PRL-2001,Hong-PRB-2009,Eom-PRL-2006,Wang-PRL-2006,Hsu-SS-2010}, peculiarities of the growth of metal nanoislands \cite{Hsu-SS-2010,Fokin-PSS-2010,Jiang-PRB-2004}, vortex states in superconducting nanostructures \cite{Cren-PRL-09,Moore-SST-2015}, and electronic properties of superconductor--normal metal hybrid structures \cite{Cherkez-PRX-2014}. The main methods for studying electronic states in Pb films are low-temperature scanning tunneling microscopy (STM) and spectroscopy  \cite{Altfeder-PRL-1997,Altfeder-PRL-2002,Su-PRL-2001,Hong-PRB-2009,Eom-PRL-2006,Wang-PRL-2006,Hsu-SS-2010,Fokin-PSS-2010,Jiang-PRB-2004,Cren-PRL-09,Moore-SST-2015,Cherkez-PRX-2014},  transport measurements  \cite{Jalochowski-PRB-1988,Miyata-PRB-2008}, and photoemission  studies \cite{Miyata-PRB-2008,Dil-PRB-2006,Mans-PRB-2002,Mulin-RPP-2002,Ricci-PRB-2009,Slomski-PRB-2011}.


Peaks of the differential tunneling conductivity were detected for some values of the potential $U^{\,}_n$ of ultrathin Pb film and island samples by tunneling spectroscopy  methods \cite{Altfeder-PRL-1997,Hong-PRB-2009,Eom-PRL-2006,Wang-PRL-2006,Su-PRL-2001,Hsu-SS-2010,Moore-SST-2015}. A correlation between the spectrum of the $U^{\,}_n$ values and the local thickness of the Pb layer was found and interpreted in terms of the resonant tunneling of electrons through quantum-confined levels in a quasi-two-dimensional electron gas. The energy of quantum-confined levels for electrons in a one-dimensional  potential well is determined by the Bohr--Sommerfeld quantization rule \cite{Altfeder-PRL-2002}
    \begin{equation}
    \varphi^{\,}_1 + \varphi^{\,}_2 + 2k^{\,}_{\perp,n} D = 2\pi n,
    \label{BohrSommerfeld}
    \end{equation}
where $\varphi^{\,}_{1}$ and $\varphi^{\,}_{2}$ are the phase shifts of the electronic wave reflected from the metal--vacuum and metal--substrate interfaces, respectively; $k^{\,}_{\perp,n}$  is the spectrum of allowed values of the wave vector transverse with respect to the interfaces; $D$ is the thickness of the layer; and $n$ is the number of nodes of a standing electron wave.

The authors of \cite{Altfeder-PRL-2002,Mans-PRB-2002,Mulin-RPP-2002,Ricci-PRB-2009} tried to extract the phases $\varphi^{\,}_{1}$ and $\varphi^{\,}_{2}$ from tunneling and photoemission spectroscopy data. The authors of \cite{Altfeder-PRL-2002} demonstrated the possibility of visualizing structure of atoms of the lower interface under the metal layer by the scanning tunneling microscopy due to the dependence of the phase shift $\varphi^{\,}_{2}$ on the lateral coordinates. Quantum--confined states and the corresponding features of the tunneling conductivity or optical properties were also revealed in Ag and Cu films \cite{Mulin-RPP-2002,Chiang-PRB-2011} and In islands \cite{Dil-PRB-2006,Altfeder-PRL-2004}.

Local tunneling spectroscopy \cite{Altfeder-PRL-1997,Hong-PRB-2009,Eom-PRL-2006,Wang-PRL-2006,Su-PRL-2001,Hsu-SS-2010,Moore-SST-2015} performed for a limited number of points cannot reliably determine the boundaries of regions with a constant thickness of the Pb layer. As far as we know, the application of scanning bias--modulation spectroscopy, which involves simultaneous  acquisition of a topographic image and a map of the density of states for a given energy, for ultrathin Pb films has not yet been discussed.

\begin{figure*}[ht!]
\centering{\includegraphics[width=14.2cm]{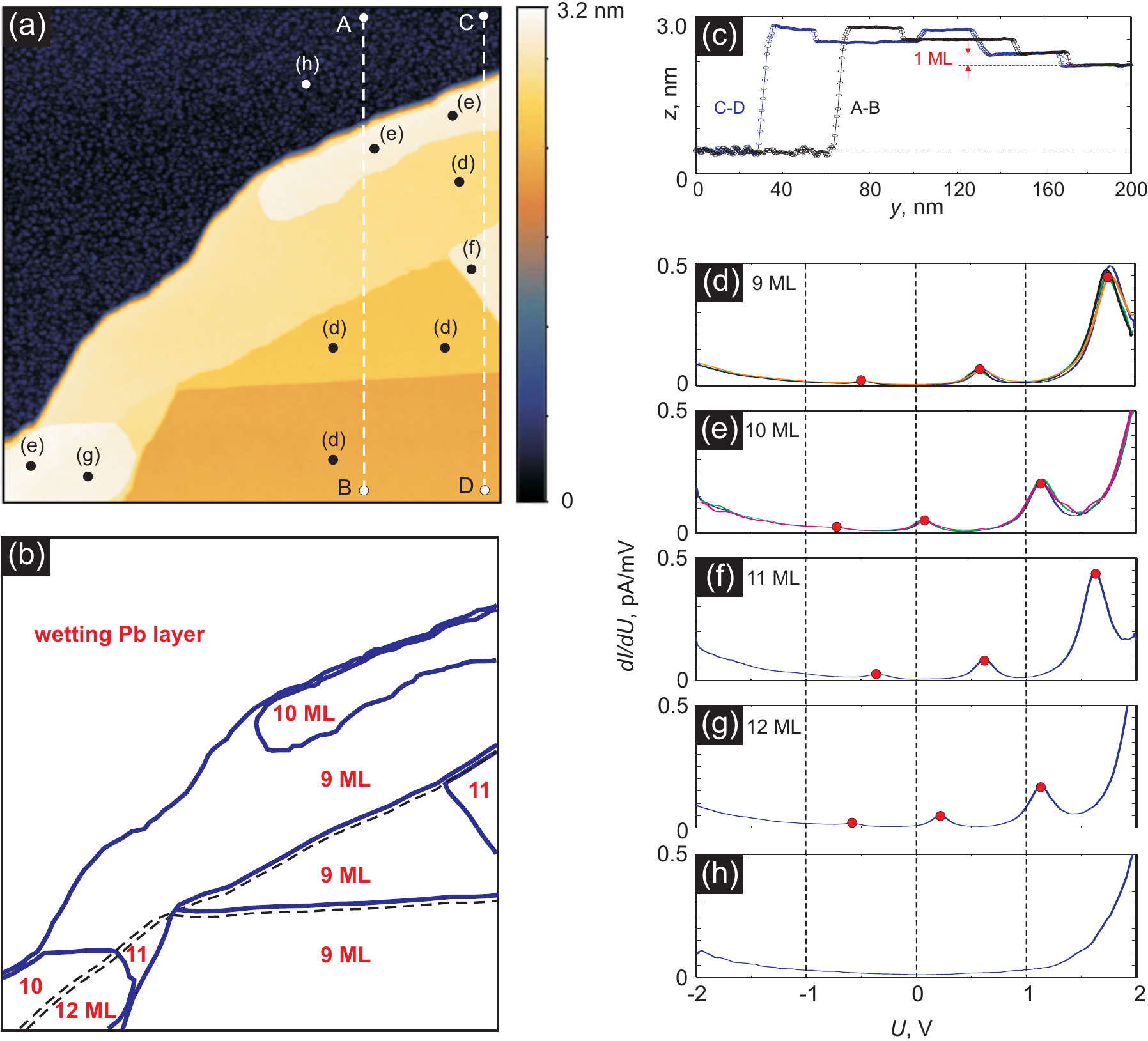}}
\caption{{\bf (a)} Scanning tunneling microscopy image 230$\times$230~nm$^2$ of the surface of the Pb island on top of Si(111)$7\times7$ obtained at $U=+2.00$\,V and $I=50$\,pA. {\bf (b)} Schematic diagram of the structure of the studied area. The thick solid lines indicate the boundaries of the terraces at the upper interface of the Pb film. The dashed lines show the steps of monoatomic height in the Si(111) substrate. Numbers indicate the nominal thickness measured from the level of the wetting layer and expressed in units of $d^{\,}_{ML}$. {\bf (c)} Profiles of the surface along the A--B and C--D lines. {\bf (d--h)} Differential conductivity $dI/dU$ versus $U$ at several points of the surface marked in panel (a) with the thicknesses of the Pb layer of 9 ML (d), 10 ML (e), 11 ML (f), 12 ML (g), and within the wetting layer (h); the measurements were performed under the initial condition at $U=+2.00$\,V and $I=200$\,pA.}
\label{Fig1}
\end{figure*}

In this work we analyze the possibility of determining the thickness of Pb films and visualizing features of the  microstructure of the substrate (steps of monoatomic height and foreign inclusions) under the Pb layer by low--temperature bias--modulation scanning tunneling spectroscopy.

\section{Experimental procedure}

The preparation of the surface of substrates, thermal deposition of Pb and the investigation of the electrophysical properties of Pb nanostructures is performed on a UHV LT SPM Omicron Nanotechnology ultrahigh--vacuum setup. The topography of prepared  structures is studied by scanning tunneling microscopy at a temperature of 78 K in the regime of a given tunneling current at a constant potential $U$ of the sample with respect to the tip of a tunneling microscope. Etched tungsten wires with apex cleaned by electron bombardment in ultrahigh vacuum are used as tips. The quality of prepared tips was tested by scanning Si(111)$7\times7$ and Au(111) surfaces. The electronic properties of Pb nanostructures are studied by scanning tunneling spectroscopy, which involves the measurement of local current--voltage ($I-U$) characteristics of a tip--sample tunnel junction or series of such characteristics at a fixed position of the tip. Furthermore, bias--modulation tunneling spectroscopy provides maps of local differential conductivity $dI/dU$ as functions of the coordinates for a given average bias voltage by the filtration of the oscillating component of $I$ at the modulation frequency of the tip potential $f_0=11.111$~kHz using a Stanford Research SR 830 lock--in amplifier.

The thermal deposition of Pb was performed in two stages. First, lead (Alfa Aesar, purity of 99.99\%) 
was deposited  from a Mo crucible by  an EFM3 electron--beam evaporator on a preliminarily prepared Si(111)$7\times7$  surface at a pressure of $2\times10^{-10}$ mbar at a rate of about 0.01 nm/min at room temperature for 6 min.  Such a procedure ensured the formation of the amorphous  wetting Pb layer without islands, which was confirmed by the subsequent scanning tunneling microscopy analysis. Then, to obtain  two--dimensional Pb islands, lead was deposited on the wetting layer at a pressure of $6\times10^{-10}$ mbar at a rate of about 0.5 nm/min at room temperature for 4 min, which corresponds to the average thickness of the deposited layer of 6 ML of lead.

\section{Results and discussion}

Figure~\ref{Fig1}a  shows a typical topographic image of the  Pb/Si(111)$7\times7$ surface aligned with  respect to the region with the lowest height. It is well known \cite{Altfeder-PRL-1997,Hong-PRB-2009,Eom-PRL-2006,Wang-PRL-2006} that the growth of Pb nanostructures occurs through the Stranski--Krastanov mechanism, which results in the formation of ultrathin islands with typical lateral dimensions from several tens to hundreds of nanometers. The A--B and C--D profiles shown in Fig.~\ref{Fig1}c indicate a quantized change in the height; consequently, the minimal change in the height of terraces should be associated with the thickness of a monolayer (ML) of lead atoms: $d^{\,}_{ML}=0.28\pm 0.01$\,nm. The estimate $d^{\,}_{ML}$ for our films coincides with the distance between atomic planes for single-crystal lead in the (111) direction: $d^{\,}_{ML}=a/\sqrt{3}=0.285$\,nm, where  $a=0.495$\,nm is the lattice constant. Since all heights in Figs.\ref{Fig1}a--\ref{Fig1}c are measured from the amorphous wetting layer \cite{Altfeder-PRL-1997}, the actual thickness of the film $D$ is the sum of the parameter $d^{\,}_w$, which is determined by the thickness of the wetting layer and a finite radius of localization of wave functions beyond the layer, and the nominal thickness $d$ of islands on the wetting layer. The available spectrum of heights in Fig.~\ref{Fig1}a can be expressed in terms of the number of monolayers $N=d/d^{\,}_{ML}$ and the following schematic representation can be proposed (see Fig.~\ref{Fig1}b).

\begin{figure*}
\centering{\includegraphics[width=16cm]{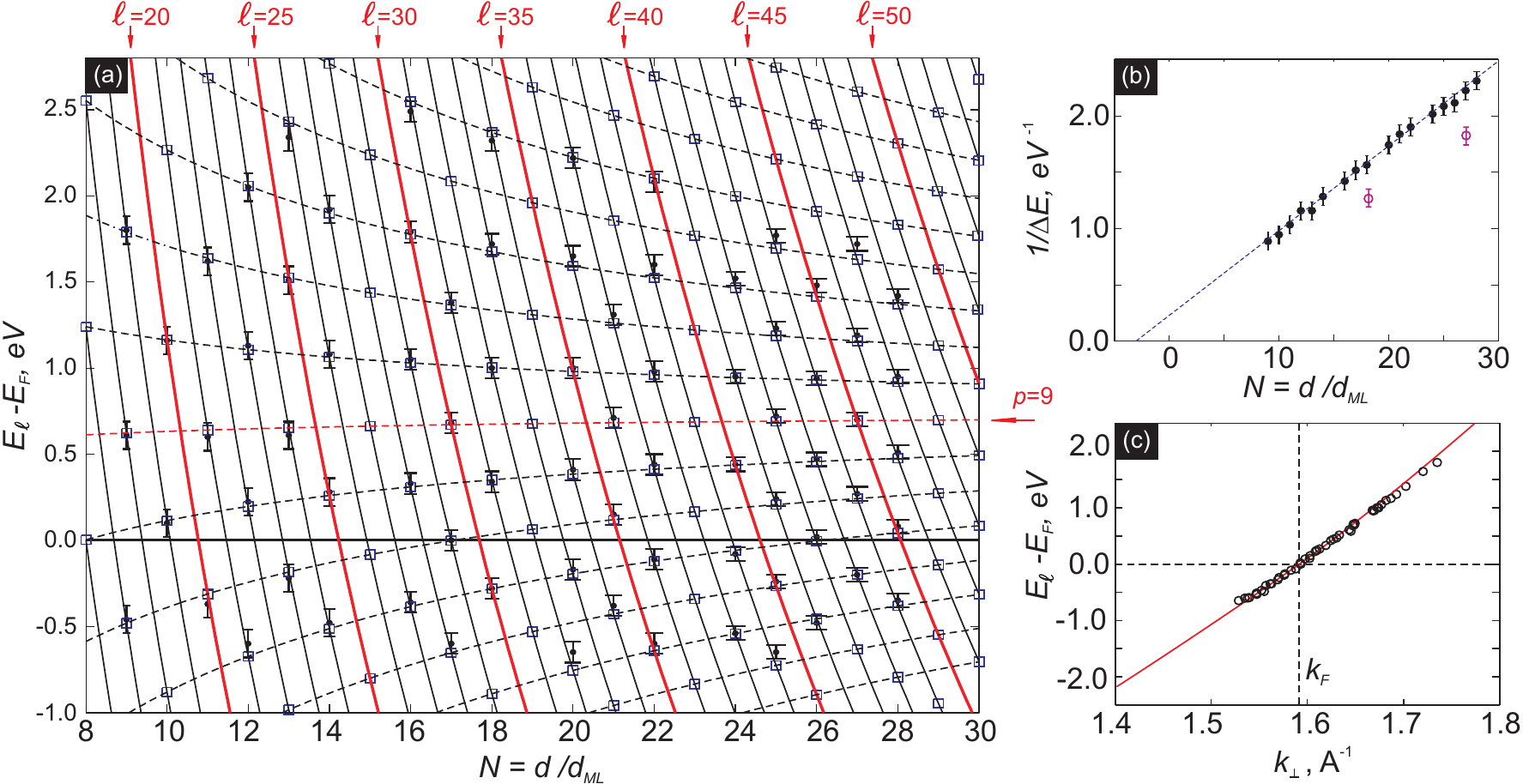}}
\caption{{\bf (a)} Positions of the maxima $E^{\,}_n$ of the tunneling conductivity versus the nominal thickness of the layer $N=d/d^{\,}_{ML}$ for the Pb/Si(111)$7\times7$ system. The symbols $\Box$ mark the expected positions of the quantum--confined levels in the model specified by Eq.~(\ref{Eq2}) taking into account that $D=d^{\,}_{ML}\,(N+3)+0.02$~nm \cite{Su-PRL-2001}. The thin and thick solid lines correspond to constant principal quantum numbers $\ell$ and the dashed lines correspond to integer values of the parameter $p=2\ell-3N$. {\bf (b)}  Difference $\Delta E = E^{\,}_{\ell}-E^{\,}_{\ell-1}$ (for energies near the Fermi level) versus $N$ for Pb/Si(111)$7\times7$ ($\bullet$) and Pb/HOPG ($\circ$). The linear extrapolation of the dependence $1/\Delta E$ on the thickness of the layer makes it possible to estimate the thickness of the wetting layer for the Pb/Si(111)$7\times7$ system as $d^{\,}_w\simeq 3d^{\,}_{ML}$. {\bf (c)}  Spectrum $E^{\,}_{\ell}(k^{\,}_{\perp,\ell})$  reconstructed from the measurements shown in the panel (a); $k_F\simeq 1.593$~\AA$^{-1}$ in the extended band scheme. The solid line corresponds to the spectrum $E(k)$ of bulk Pb in the (111) direction \cite{Pap}.}
\label{Fig2}
\end{figure*}

Figures~\ref{Fig1}d--\ref{Fig1}h show the dependences of $dI/dU$ on $U$ obtained for the Pb/Si(111)$7\times7$ system at various points. It was found that tunneling spectra at various points of the surface with the same thickness are identical (see Figs.~\ref{Fig1}d--\ref{Fig1}e). An important feature of the resulting spectra is the presence of almost equidistant peaks of the conductivity; the positions of these peaks depend on the local thickness of the layer \cite{Altfeder-PRL-1997,Hong-PRB-2009,Eom-PRL-2006,Wang-PRL-2006}. Such peaks of the conductivity are absent in local tunneling measurements within the wetting Pb layer in the range of $\pm 2\,$V.

Using the measurements on several Pb islands on Si(111)$7\times7$ with a thickness of 2.5 nm (9 ML) to 14.3 nm (50 ML), we composed the $E^{\,}_m-N$ diagram indicating the positions of maxima of the conductivity versus the nominal thickness of the Pb layer expressed in units of $d^{\,}_{ML}$ (Fig.~\ref{Fig2}a). An  interval  between  two resolved quasistationary states $\Delta E=E^{\,}_{\ell}-E^{\,}_{\ell-1}$  near the Fermi level $E^{\,}_F$ can be estimated. It is easy to see that the dependence of $\Delta E^{-1}$ on $N$ can be approximated with a good accuracy by a linear function: $\Delta E^{-1}\simeq A\,(N+3)$, where the parameter $A=0.0753\,$eV$^{-1}$ was determined by the least squares approximation;  consequently, $d^{\,}_{w}\simeq 3d^{\,}_{ML}$ \cite{Altfeder-PRL-1997}. The  interval  between  quantum--confined levels for the wetting layer should be $\Delta E=(3A)^{-1}=4.5\,$eV, which explains the absence of pronounced  peaks  of  the  tunneling  conductivity  in the considered range of bias voltages. Preliminary data obtained for the Pb/HOPG system indicate that the parameters $d^{\,}_w$ and $v^{\,}_F$ can depend on the material of the substrate. Therefore, the $E^{\,}_m-N$ diagram (see Fig.~\ref{Fig2}a) is not universal and is sensitive to the material of the substrate.

The obtained data can be interpreted within the simplest model of localization of a particle with the effective mass $m^*$ in a one-dimensional potential well with the width $D$ and infinite walls. In this case, $\varphi^{\,}_{1,2}=\pi/2$ and solutions of Eq.~(\ref{BohrSommerfeld}) are states with $k^{\,}_{\perp,\ell}=\pi \ell/D$, where $\ell=1, 2, \ldots$ is the number of electronic half--waves. Then,
    \begin{equation}
    E^{\,}_{\ell} = E^{\,}_0 + \frac{\hbar^2k^{2}_{\perp,\ell}}{2m^*} \simeq E^{\,}_F + \hbar v^{\,}_F\,\left(\frac{\pi \ell}{D} - k^{\,}_F\right),
    \label{Eq2}
    \end{equation}
where $E^{\,}_0 = E^{\,}_F - \hbar^2k^{2}_{F}/2m^*$ is the bottom of the conduction band and $k^{\,}_F$ and $v^{\,}_F=\hbar k^{\,}_F/m^*$ are the wave vector  and semiclassical velocity at the Fermi level, respectively. Since $D=d^{\,}_{ML}\,(N+3)$ we obtain $\Delta E = \pi\hbar v^{\,}_F/D$ and $d (\Delta E)^{-1}/dN = d^{\,}_{ML}/(\pi\hbar v^{\,}_F)$, what makes it possible to determine the Fermi velocity for electrons in the Pb(111) film as $v^{\,}_F = d^{\,}_{ML}/(\pi\hbar A)\simeq 1.83\times 10^8\,$cm/s. The resulting value is in agreement with the data reported in \cite{Altfeder-PRL-1997,Su-PRL-2001}.

Expression (\ref{Eq2}) for each $m$ value predicts a hyperbolic dependence of the energy on $N$ if the number of monolayers is treated as a continuous parameter (thin and thick solid lines in Fig.~\ref{Eq2}a). It is noteworthy that one of the peaks of the conductivity for films with $N=17$ and 26 is observed at zero bias and, consequently, allowed $k^{\,}_{\perp}$ values should be close to $k^{\,}_{F}$.  Counting the number of possible hyperbolic lines between resonance states on the Fermi level for $N=17$ and 26, we find that the principal quantum numbers of these states differ from  each  other  by  13. Let $\ell^{\,}_0$ be the number of  the quantum state corresponding to the peak at zero bias for $N=17$;  then, $\ell^{\,}_0/(17+3)=(\ell^{\,}_0+13)/(26+3)$, from which $\ell^{\,}_0=29$.  All other peaks can be indexed automatically (see  Fig.~\ref{Fig2}a). The presented way of numbering of peaks gives the values coinciding with the data from other works \cite{Su-PRL-2001}. Consequently, we can estimate $k^{\,}_F=\pi \ell^{\,}_0/[d^{\,}_{ML}\,(17+3)]=15.93\,$nm$^{-1}$ in the extended band scheme and $m^*=\hbar k^{\,}_F/v^{\,}_F=1.01\,m^{\,}_0$, where $m^{\,}_0=9.1\times 10^{-31}\,$kg is the mass of the free electron. Furthermore, it is easy to transform the $E^{\,}_{\ell}-N$ diagram to the $E^{\,}_{\ell}(k^{\,}_{\perp,{\ell}})$ dependence  (see Fig.~\ref{Fig2}c). As expected, the dependence of $E^{\,}_{\ell}$ on $k^{\,}_{\perp,\ell}$ at $E\simeq E^{\,}_F$ is close to a linear function, which is in agreement with experimental data and calculations for bulk Pb \cite{Hong-PRB-2009,Pap}.

The  Fermi wavelength can be estimated as $\lambda^{\,}_{F}=2\pi/k^{\,}_{F}=0.394\,$nm; therefore, the ratio $\lambda^{\,}_{F}/d^{\,}_{ML}$ close to 4/3 \cite{Su-PRL-2001}. Consequently, the energy of an electronic state with the index $\ell$ (i.e., the number of half-waves) near $E^{\,}_F$  for the film with a thickness of $N$ ML should be close to the energy of the state with the index $\ell+3$ for the film with a thickness of $(N+2)$ ML. Using Eq.~(\ref{Eq2}), we  write the energy of states for the integer index $p=2\ell-3N$  in the form
    \begin{equation}
    E^{\,}_p \simeq E^{\,}_F + \hbar v^{\,}_F\,\left(\frac{3}{2}\,\frac{\pi}{d^{\,}_{ML}}\,\frac{(N+p/3)}{(N+3)} - k^{\,}_F\right),
    \label{Eq3}
    \end{equation}
The dependences of $E^{\,}_p$ on $N$ for various $p$ values are shown in Fig.~\ref{Fig2}a. It is remarkable that the energy of states with $p=9$ in our model is independent of the thickness:
    \begin{equation}
    E^{\,}_{p=9} - E^{\,}_F \simeq  \frac{\hbar^2 k^{\,}_F}{m^*}\,\left(\frac{3}{2}\,\frac{\pi}{d^{\,}_{ML}} - k^{\,}_F\right),
    \label{Eq4}
    \end{equation}
The positions of peaks of the conductivity in the range of 0.6 to 0.7 eV that are observed for films with odd $N$ values are in good agreement with the estimate (\ref{Eq4}). A small experimentally observed slope of the dependence of $E^{\,}_{p=9}$ on $N$ indicates that the parameter $d^{\,}_w$, which is determined by the thickness of the wetting layer and a finite localization radius of wave functions beyond the Pb film, is not precisely equal to $3d^{\,}_{ML}$  and the relation $\lambda^{\,}_F/d^{\,}_{ML}=4/3$ is valid approximately.

\begin{figure*}
\centering{\includegraphics[width=12cm]{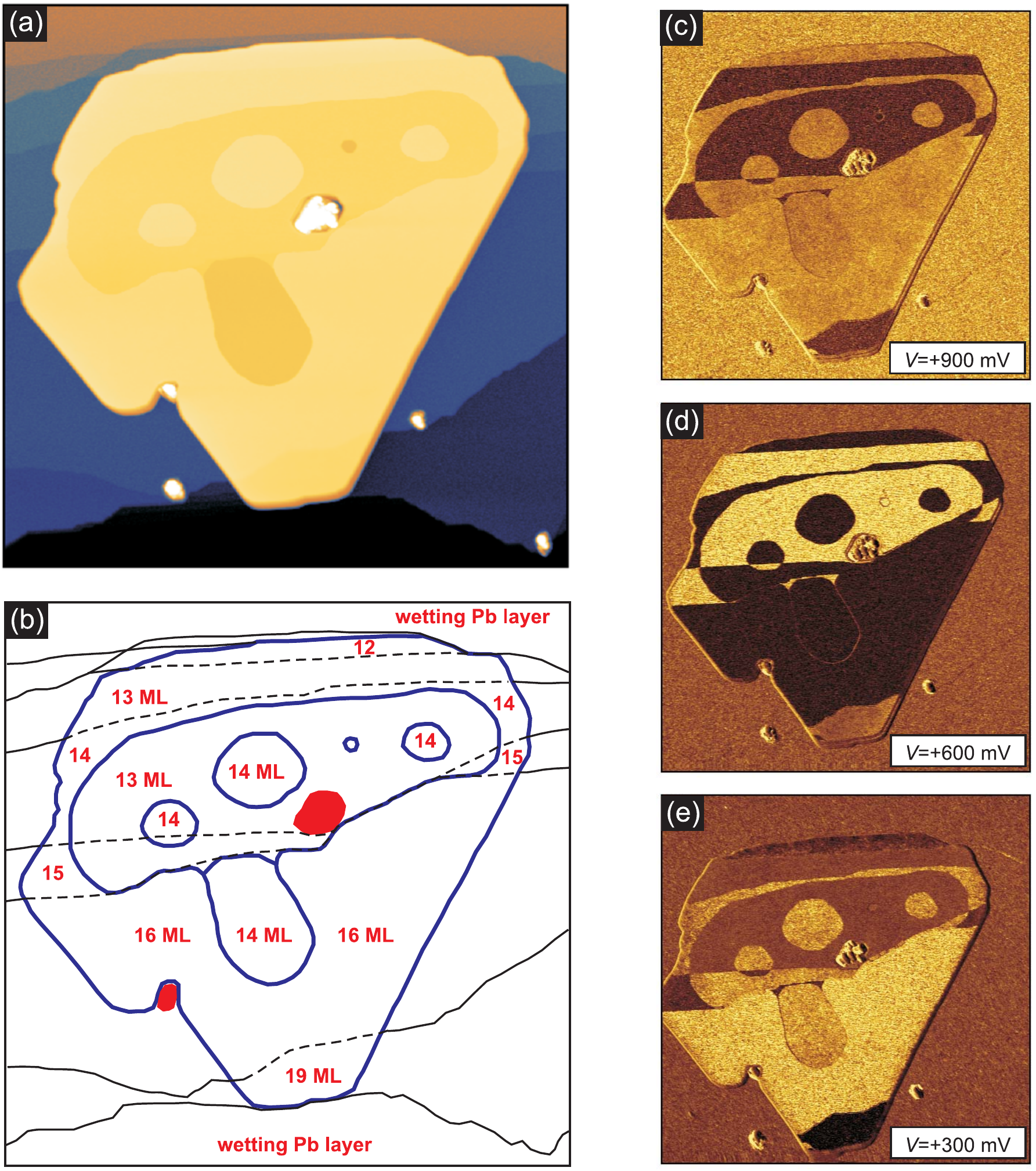}}
\caption{{\bf (a)} Scanning tunneling microscopy image 690$\times$ 690\,nm$^2$ of the Pb/Si(111)$7\times7$ surface obtained at and $I=300~$pA. {\bf (b)} Schematic of the structure of the studied area. The thick solid and dashed lines show monatomic-height steps in the Si(111) substrate. Numbers indicate the nominal thickness in terms of $d^{\,}_{ML}$. {\bf (c--e)} Maps of the differential conductivity (the average tunneling current is $I=300~$ pA, the modulation amplitude of the tip potential is 50 mV, and the frequency is 11.111 kHz) at bias voltages of (c), (d) and (e); brighter (darker) regions correspond to higher (lower) tunneling conductivity. }
\label{Fig3}
\end{figure*}

\begin{figure}[th!]
\centering{\includegraphics[width=5.5cm]{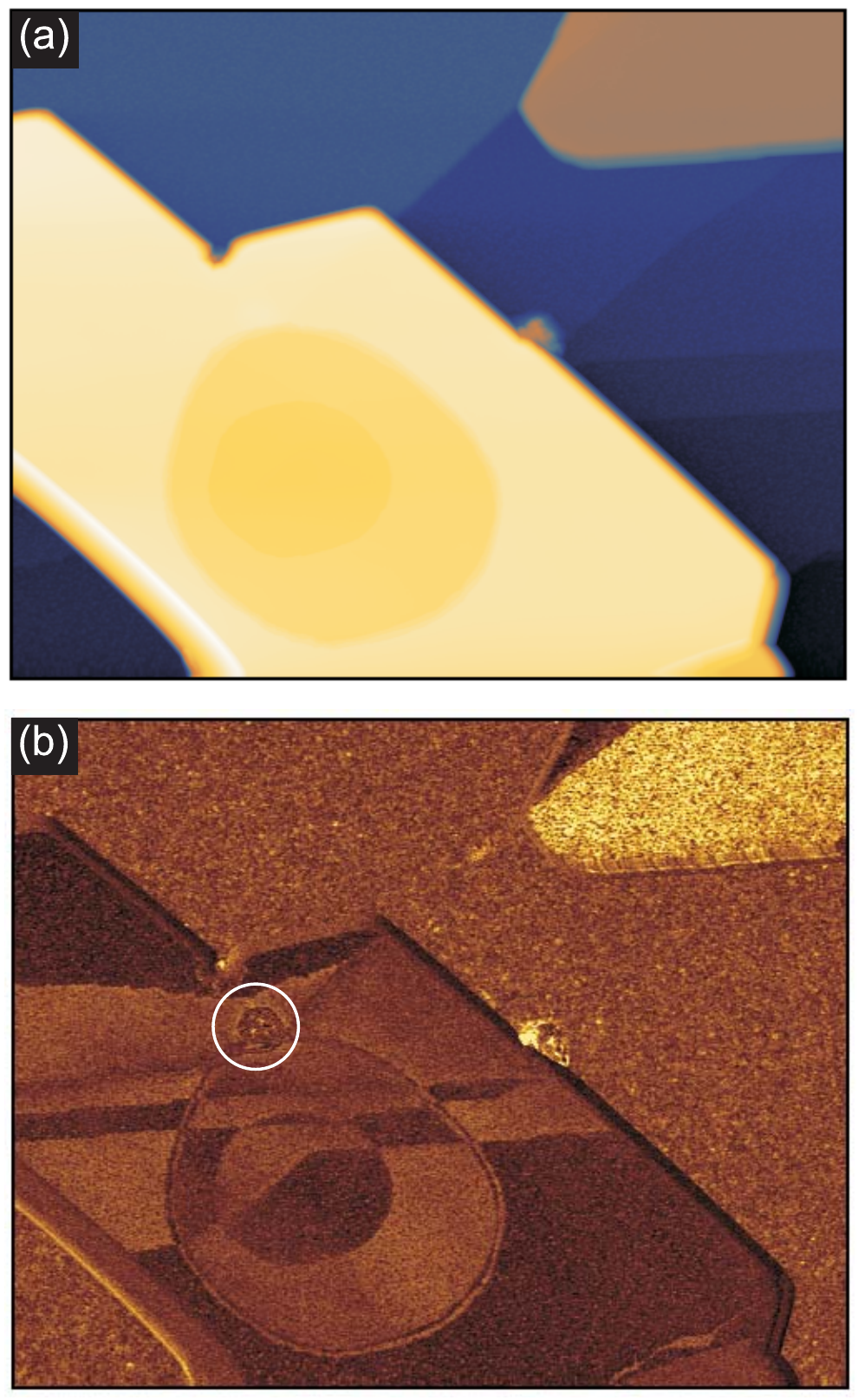}}
\caption{{\bf (a)} Scanning tunneling microscopy image 450$\times$350~nm$^2$ of the Pb/Si(111)7$\times$7 surface obtained at $U=+0.30~$V and $I=200$~pA. (b) Map of the differential conductivity ($U=+0.30~$V and $I=200$~pA, modulation amplitude is 50 mV). The circle indicates a foreign inclusion that is not visible in the topography but becomes noticeable at scanning in the bias--modulation mode.}
\label{Fig4}
\end{figure}

\begin{figure}[th!]
\centering{\includegraphics[width=6cm]{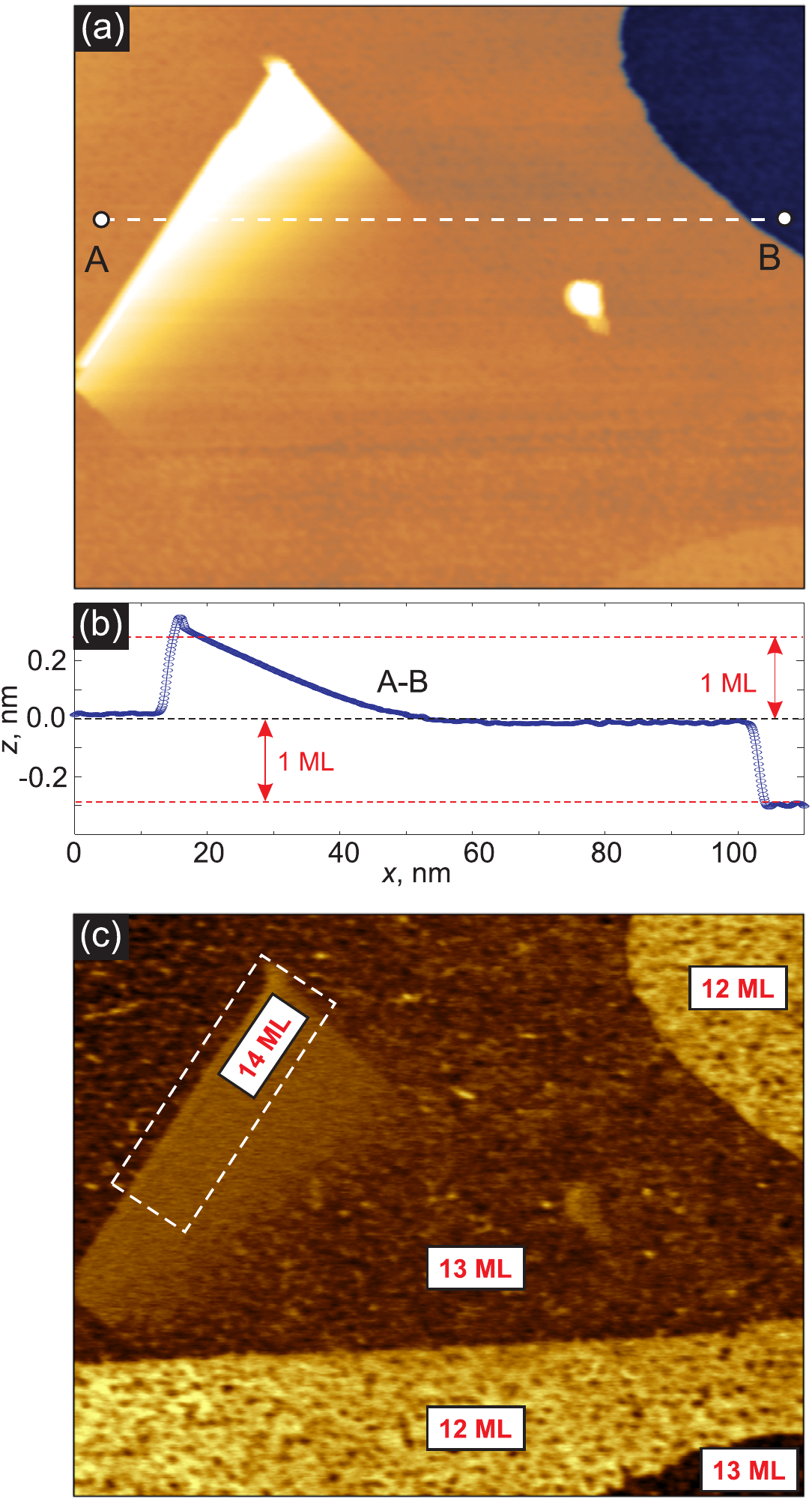}}
\caption{{\bf (a)} Scanning tunneling microscopy image 115$\times$92 nm$^2$ of the surface Pb/Si(111)7$\times$7 acquired for $U=+0.20$~V and $I=200$~pA. {\bf (b)} The profile along the A--B line. {\bf (c)} Map of local differential tunneling conductivity $dI/dU$ at $U=+0.20$\,V, $I=200$\,pA and the amplitude of the bias modulation 50 mV; the labels indicate the local thickness of the Pb films with respect to the wetting Pb layer.}
\label{Fig5}
\end{figure}

Figure~\ref{Fig3} shows the results of bias--modulation scanning tunneling spectroscopy study of the topography and the local differential conductivity of Pb nanoislands.  The  comparison  of  the  topographic  image (Fig.~\ref{Fig3}a) with maps of the local density of states at various energies (Figs.~\ref{Fig3}d--\ref{Fig3}f) clearly shows that regions with an identical thickness of the Pb layer correspond to regions of an equal intensity on differential conductivity maps. This correspondence makes it possible to identify terraces of Pb islands with the same thickness even when an island is located in a complicated system of monatomic steps in the substrate or when the scanning region includes a part of the island with the surrounding wetting layer (Figs.~\ref{Fig3} and \ref{Fig4}). In particular, according to the diagram shown in Fig.~\ref{Fig2}a, a measurement at $U=+0.6\,$V makes it possible to identify terraces with an odd number of monolayers with respect to the wetting layer because one of the conductivity  peaks for such terraces lies near +0.6\,V (see Fig.~\ref{Fig3}d). A single differential conductivity map is obviously insufficient to reconstruct the thicknesses of all regions, but several maps recorded at different biases significantly simplify the interpretation of a topographic image. It is noteworthy that bias--modulation scanning tunneling spectroscopy allows to reveal hidden details of the image (e.g., steps in the substrate and defects), which are completely covered with a metal layer and, hence, are invisible on the topographic image. For example, an invisible cluster of a different material is revealed under a Pb island with the atomically smooth surface in Fig.~\ref{Fig4}b.

We would like to illustrate a procedure of the determination of the local thickness of a Pb film based on a single map of the differential tunneling conductance. The topographic image in Fig.~\ref{Fig5}a indicates that there are one developed monoatomic step and one appearing monoatomic step, caused by two screw dislocations, on the top surface of the Pb film. The map of the local differential conductance shown in Fig. 1c points to the additional monoatomic steps on the bottom surface of the Pb film. Thus, one can conclude that there are three regions of different nominal thickness: $N_0$ (bright areas in Fig.~\ref{Fig5}c), $N_0+1$ (dark areas in Fig.~\ref{Fig5}c) and close to $N_0+2$ (area of the intermediate intensity in Fig.~\ref{Fig5}c). Taking into account the diagram in Fig.~\ref{Fig2}a, one can conclude that the peak of tunneling conductance at +0.2 V should correspond to the Pb film with $N_0=12$. As a consequence, the bright and dark areas in the $dI/dU$ map should be attributed to $N=12$ and $N=13$, respectively; while the thickness for the area of the intermediate intensity within the appearing monoatomic step varies from $N=13$ to $N=14$ (see labels in Fig.~\ref{Fig5}c). This conclusion was supported by local STS measurements, which are similar to that shown in Figs.\ref{Fig1}f--g.

\section{Conclusion}

Pb nanoislands grown on the Si(111)$7\times7$ surface have been experimentally studied by low--temperature scanning tunneling microscopy and bias--modulation scanning tunneling spectroscopy. It has been shown that the local differential tunneling conductivity include pronounced peaks of the conductivity. The energies of the corresponding quasistationary states depend on the thickness of the Pb layer at the location of the tip of a scanning tunneling microscope. A method for indexing peaks of the conductivity has been proposed on the basis of the unambiguous determination of quantum numbers for states at the Fermi level. Within the simplest model of a particle in a potential well with infinite walls, the resonance energy of the conductivity peak, which is almost independent of the film thickness, has been calculated. The relation of the energy of this peak to the microscopic parameters of the film ($d^{\,}_{ML}$, $k^{\,}_F$, and $m^*$) has been discussed. It has been shown that bias--modulation scanning tunneling spectroscopy at liquid nitrogen temperatures allows visualizing hidden defects under the metal layer with a thickness of 15 nm with a subnanometer spatial resolution. In particular, monatomic steps in the substrate, as well as foreign inclusions under the metal layer,  which are not manifested in a topographic image, have been revealed.

\section{Acknowlednements}

We are grateful to V.\,S. Stolyarov and S.\,I. Bozhko for valuable remarks and assistance in the work. The work was performed with the use of equipment at the Common Research Center 'Physics and Technology of Micro- and Nanostructures' at Institute for Physics of  Microstructures,  Russian Academy of Sciences (Nizhny Novgorod, Russia). The work of S.\,S.\,U. and A.\,Yu.\,A. was supported by the Russian Foundation for Basic Research (project nos. 15-42-02416 and 16-02-00727). The work of A.\,V.\,P. was supported by the Russian Science Foundation (project no. 15-12-10020).

\end{document}